\documentclass[aip, jmp, reprint]{revtex4-1}
\usepackage{graphicx}
\usepackage{amsmath}
\usepackage{amssymb}
\usepackage{dcolumn}
\usepackage{bm}
\usepackage[retainorgcmds]{IEEEtrantools}
\usepackage{epstopdf}
\usepackage{color}

\allowdisplaybreaks
\graphicspath{ {./Images/} }
\begin{document}

\title{Do micromagnets expose spin qubits to charge and Johnson noise?}

\author{Allen Kha}
\affiliation{School of Physics, The University of New South Wales, Sydney
NSW 2052, Australia}

\author{Robert Joynt}
\affiliation{Department of Physics, University of Wisconsin, Madison, Wisconsin 53706, USA}
\affiliation{School of Physics, The University of New South Wales, Sydney
NSW 2052, Australia}

\author{Dimitrie Culcer}
\affiliation{School of Physics, The University of New South Wales, Sydney
NSW 2052, Australia} 

\date{\today}

\begin{abstract}
An ideal quantum dot spin qubit architecture requires a local magnetic field for one-qubit rotations. 
Such an inhomogeneous magnetic field, which could be implemented via a micromagnet, couples the qubit subspace with background charge fluctuations causing dephasing of spin qubits. In addition, a micromagnet generates magnetic field evanescent-wave Johnson noise. 
We derive an effective Hamiltonian for the combined effect of a slanting magnetic field and charge noise on a single-spin qubit and estimate the free induction decay dephasing times $T_{2}^{\ast}$ for Si and GaAs. 
The effect of the micromagnet on Si qubits is comparable in size to that of spin-orbit coupling at an applied field of $B=1$T, whilst dephasing in GaAs is expected to be dominated by spin-orbit coupling. 
Tailoring the magnetic field gradient can efficiently reduce $T_{2}^{\ast }$ in Si. 
In contrast, the Johnson noise generated by a micromagnet will only be important for highly coherent spin qubits.
\end{abstract}

\preprint{AIP/123-QED}

\maketitle

Solid state spin systems, in which coherence times of up to a few seconds have been measured, \cite{Tyryshkin,Amasha_PRL2008,BarGill} are promising candidates for scalable quantum computer architectures, \cite{Loss_PRA98,Burkard_PRB99}with silicon ideal for hosting spin-based qubits. \cite{Huang_PRB2014,Itoh_MRS,Zwanenburg1,Liu_PRB2008,Prada_PRB08,Gamble_PRB2012} 
Addressing individual qubits is vital, yet using electron spin resonance requires bulky on-chip coils that dissipate heat close to the electron. \cite{natKoppens,Kawakami2013} 
However, electric fields can be generated locally with small, low voltage electrodes, and electrical spin rotations \cite{Nowack_Science07,Pioro-L2} can be accomplished by the modulation of a quantum dot electric field in a slanting static magnetic field. \cite{Pioro-L1,Pioro-L2,PhysRevLett.96.047202,Kawakami2014}

Quantum dot spin qubits are typically located near semiconductor interfaces where defects are present.\cite{vanderZiel, Jung, Hitachi, Helms,Machlin2006215} 
The resulting fluctuations in the local electric field are a well known source of dephasing in charge qubits, \cite{Paladino1,Nakamura_PRL2002,PhysRevLett.110.146804,PhysRevLett.110.136802,Petersson_PRL10,Hayashi_DQD_PRL03} as well as relaxation \cite{PhysRevB.89.195302} and dephasing of spin qubits\cite{Berme} when spin-orbit coupling is present. 
Here we show that the micromagnet couples spin qubits to charge fluctuations and causes dephasing even when spin-orbit coupling is absent. 
In addition, a ferromagnet contains currents and spins that fluctuate due to both thermal and quantum effects. 
This generates random magnetic fields nearby. 
Thus a micromagnet in the vicinity of spin qubits can cause spin dephasing and relaxation.
This effect is similar to the relaxation caused by evanescent-wave Johnson noise\cite{LangsjoenPhysRevB.89.115401} recently observed in NV centers in diamond close to metallic surfaces. \cite{Kolkowitz06032015} 
In the case of a micromagnet, however, we must consider the dissipative magnetic, not electrical, response of the noise source. 
An analysis of this kind has recently been done in Ref. \onlinecite{Neumann_JAP2015} for one type of micromagnet design. 
Although we treat here a different design, our analytical results are in qualitative agreement with the numerical results of Ref. \onlinecite{Neumann_JAP2015}.

In this paper we study two effects: qubit dephasing in the presence of (A) the combined effects of charge noise and an inhomogeneous magnetic field, and (B) Johnson-type magnetic field noise generated by a micromagnet. 
We model the first effect as random telegraph noise (RTN) and $1/f$ noise\cite{PhysRevB.73.180201,Paladino_1/f_RMP14,Burkard_Decoh_AdvPhys08,Culcer_APL09,Clarke_Nat2008,Makhlin_RMP2001} together with an inhomogeneous magnetic field. 
We compare dephasing times $T_{2}^{\ast }$ for identical dot designs in Si and GaAs. 
For the second effect we derive the appropriate formulas that govern the strength of the fluctuating magnetic fields in the vicinity of the micromagnet and their effect on the qubit. 
For the parameters appropriate to a representative device architecture this effect is not large. It will be important as coherence times approach the range of $1-10$ s.

\textit{Charge noise combined with an inhomogeneous magnetic field}. 
We focus on a single-spin qubit implemented in a symmetric gate defined quantum dot, located at the flat interface of a semiconductor heterostructure (Si/SiGe or GaAs/AlGaAs). 
The two-dimensional electron gas (2DEG) lies in the $xy$-plane. 
The gate confinement is assumed harmonic $V(x,y)=\tfrac{1}{2} m^{\ast}\omega^{2}(x^{2}+y^{2})$, where $m^{\ast}$ is the in-plane effective mass, $\omega=\frac{\hbar }{m^\ast a^{2}}$ the oscillator frequency and $a$ the dot radius. 
The Hamiltonian for the electron kinetic energy and confinement in the $xy$-plane
	\begin{IEEEeqnarray}{rCl}
	H_0&=&-\frac{\hbar^2}{2m^*}\bigg(\frac{\partial^2}{\partial x^2}+\frac{\partial^2}{\partial y^2}\bigg)+\frac{\hbar^2}{2m^*a^4}(x^2+y^2).
	\end{IEEEeqnarray}
For the ground state $D_0(x,y) = (1/a\sqrt{\pi}) \, e^{-\frac{x^2+y^2}{2a^2}}$ with energy $\varepsilon _{0}=\frac{\hbar^{2}}{2m^{\ast }a^{2}}$. 
For the twofold degenerate first excited state $D_{\pm}(x,y)=(1/a^2\sqrt{\pi})\, (x\pm iy)e^{-\frac{x^2+y^2}{2a^2}}$ and $\varepsilon_{1} = 3\varepsilon_0$.
We initially model charge noise by a single charge trap located in the $xy$-plane at a distance $x=x_d$ from the dot center. 
The noise Hamiltonian $H_N(t)$ is a random function of time and represents a fluctuating Coulomb interaction between the electron in the dot and one at $x_d$. 
In the absence of screening the time-independent Coulomb potential $V_C(x,y)=\frac{e^2}{4\pi\varepsilon_0\varepsilon_r r_d}$, where $\varepsilon_0$ is the
permittivity of free space, $\varepsilon_r$ the dielectric constant and $r_d=\sqrt{(x-x_d)^2+y^2}$ the distance between the defect and the dot center.
The non-zero matrix elements in $H_N$ are $v_0=\langle D_0|V_C|D_0\rangle$, $v_1=\langle D_\pm|V_C|D_\pm\rangle$, $\alpha=\langle D_0|V_C|D_\pm\rangle$ and $\beta=\langle D_\pm|V_C|D_\mp\rangle$. 
Electrons in the 2DEG screen the defect potential.\cite{Supplement} 
We use the same notation to denote the matrix elements $v_0$, $v_1$, $\alpha$, $\beta$ but replace $V_C \rightarrow U_{scr}=\int d^2q/(2\pi)^2 e^{-i\mathbf{q}\cdot\mathbf{r}}U_{scr}(q)$, where $U_{scr}(q)$ is the Fourier transform of the potential\cite{Davies} and we neglect contributions from momenta $q\ge 2k_F$.

\begin{figure}[tbp]
	\centering
	\includegraphics[scale=0.5]{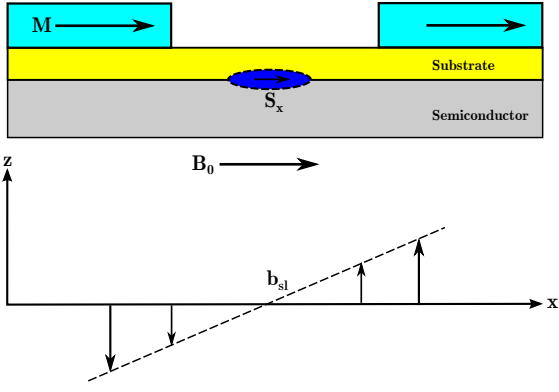}
	\caption{Sketch of the implementation of two ferromagnetic strips with
	uniform magnetisation $M$ due to the external magnetic field $B_0$ in the $x$
	-direction. The resulting out-of-plane $z$-component of the field from the
	strips has a slanting form with gradient $b_{sl}$. The spin of the electron
	is initialised $\parallel \hat{\bm x}$.}
	\label{figslanting}
	\end{figure}

We take a total magnetic field composed of a controllable homogeneous part ${\mathbf{B}}_0$ and a slanting field, \cite{Pioro-L1,Pioro-L2,PhysRevLett.96.047202} 
	\begin{IEEEeqnarray}{rCl}
	\mathbf{B}=(B_0+b_{sl}z)\boldsymbol{\hat{x}}+(b_{sl}x)\boldsymbol{\hat{z}},
	\end{IEEEeqnarray}
where the magnetic field gradient takes the value\cite{Pioro-L1,Pioro-L2} $b_{sl}=0.8$ T$\mu$m$^{-1}$. 
The magnetic field gradient is created by two ferromagnetic strips integrated on top of the QD, magnetized uniformly by the in-plane magnetic field $B_0$. 
This structure results in a stray magnetic field with an out-of-plane component that varies linearly with position with a gradient $b_{sl}$ as shown in Figure \ref{figslanting}. 
The Zeeman Hamiltonian $H_Z = \tfrac{1}{2}g\mu_B\boldsymbol{\sigma}\cdot\mathbf{B}$, where $\boldsymbol{\sigma}$ is the vector of Pauli spin matrices. 
We have six basis states $\{D_{0,\uparrow},D_{0,\downarrow},D_{+,\uparrow},D_{+,\downarrow },D_{-,\uparrow},D_{-,\downarrow }\}$, with $|\uparrow\rangle$, $|\downarrow\rangle$ the $\sigma_{x}$ eigenstates corresponding to eigenvalues $1$, $-1$ respectively. 
The qubit subspace is $\{D_{0,\uparrow},D_{0,\downarrow } \}$.

The total Hamiltonian $H = H_0+H_N+H_Z$ is given in the Supplement. \cite{Supplement} 
The magnetic field couples the qubit subspace to spin anti-aligned orbital excited states through the gradient $b_{sl}$. 
We project $H$ onto the qubit subspace by eliminating orbital excited states. \cite{Winkler} 
The resulting effective spin Hamiltonian $H_{eff}=H_{Qbt}+\tfrac{1}{2}\boldsymbol{\tilde\sigma}\cdot\mathbf{V}(t)$ expresses the perturbations of noise and the magnetic field as an effective fluctuating magnetic field in the qubit subspace (for which we use $\tilde{\bm \sigma}$) 
	\begin{IEEEeqnarray}{rCl}
	H_{eff}&=&
	\begin{pmatrix}
	\varepsilon_0^+ & 0\\
	0 & \varepsilon_0^-\\ 
	\end{pmatrix}
	+ \frac{8\eta^2\delta v\varepsilon_Z}{(\delta\varepsilon)^3}\tilde{\sigma}_z
	- \frac{4\alpha\eta\delta v}{(\delta\varepsilon)^2}\tilde{\sigma}_x
	\end{IEEEeqnarray}
where $\eta=\tfrac{1}{2}g\mu_B\langle D_0|b_{sl}x|D_+\rangle=\tfrac{1}{2}g\mu_B\langle D_-|b_{sl}x|D_0\rangle$, $\delta\varepsilon=\varepsilon_0-\varepsilon_1$, $\delta v(t)=v_0(t)-v_1(t)$ and $\varepsilon_Z = \langle D_{0,\uparrow}|H_Z|D_{0,\uparrow}\rangle$. 
The first term corresponds to a pseudospin 1/2 in an effective magnetic field assumed to be controllable, while the second arises from noise. 
The term $\propto \tilde{\sigma}_x$ is behind spin relaxation \cite{PhysRevB.89.195302} and Rabi oscillations, \cite{PhysRevB.78.195302} but its contribution to dephasing is very small as long as it is much smaller than $\delta\varepsilon$, which is true for this work. Dephasing here arises from the $\tilde{\sigma}_z$ term.

In Si particular attention must be paid to the valley splitting, which is the magnitude of the valley-orbit coupling $\Delta_v$.\cite{Culcer_PRB10} 
In this work we focus on the case $\delta\varepsilon > \Delta_v \gg\varepsilon_Z $, though we also expect our results to hold when $\Delta_v \ll\varepsilon_Z \ll \delta\varepsilon$ (the critical condition is $\delta\varepsilon > \Delta_v$). 
The inhomogeneous magnetic field will give a small matrix element coupling states from different valleys. 
For an interaction to couple valley states significantly, it must be sharp in real space \cite{Berme}. 
The $z$-dependence of the magnetic field is not sharp enough to couple valleys, and the matrix element is further diminished by the small value of the Bohr magneton. 
The small intervalley matrix elements only matter when $\Delta_v \approx \varepsilon_Z$, otherwise $\varepsilon_Z$ does not affect the valley physics. 
In this case it is known that a relaxation hotspot (a peak in $T_1$) exists,\cite{Yang_SpinVal_NC13} which suggests that $T_1$ limits $T_2^*$, and the valley degree of freedom must be taken into account explicitly, yet by adjusting the magnetic field one can tune away from this point.

The formal treatment of dephasing is summarised in the Supplement.\cite{Supplement} 
We divide the noise spectrum into two parts: (i) random telegraph fluctuators which are close to the qubit and whose effect may be resolved individually in a noise measurement and (ii) a background $1/f$ spectrum. 
We first focus on a single nearby source of RTN. 
To facilitate comparison with the spin-orbit coupling case, we consider fluctuators with shorter switching times than a cut-off of $\tau=1\,\mu$s. 
In this case $V^2\ll(\frac{\hbar}{\tau})^2$ and the initial spin decays as $\propto e^{-t/T_2^{\ast}}$, with 
	\begin{IEEEeqnarray}{rCl}
	\label{eq:T2*RTN}
	\left( \frac{1}{T_2^{\ast}}\right)_{RTN}=\frac{V^2\tau}{2\hbar^2},
	\end{IEEEeqnarray}
where for the slanting magnetic field $V=\frac{8\eta^2\delta v(t)\varepsilon_Z}{\delta\varepsilon^3}$. 
For the background $1/f$ noise we find \cite{Supplement} 
	\begin{IEEEeqnarray}{rCl}
	\label{eq:T2*1/f}
	\left( \frac{1}{T_2^{\ast}}\right)_{1/f}\approx \sqrt{\frac{\gamma_N k_BT}{2\hbar^2}}\left(\frac{8\eta^2\varepsilon_Z}{\delta\varepsilon^3}\right),
	\end{IEEEeqnarray}
where $\gamma_N$ (units of energy) is derived from experiment.

We consider sample QDs in Si/SiGe and GaAs structures and set the fluctuator switching time $\tau=1\,\mu$s, defect position $x_d=40\,$nm, Fermi wave vector $k_F=10^8\,$m$^{-1}$ and Zeeman energy $\varepsilon_Z = 60\,\mu$eV.
This does not affect $\eta$ as it is $\propto b_{sl}$ and not $\propto \mathbf{B}_0$. \footnote{With these assumptions the variations in $\mathbf{V}(t)$ arise from $\eta$, $\delta v$ and $\delta\varepsilon$.} 
We first calculate $T_{2}^{\ast}$ for a fixed quantum dot confinement energy $\delta\varepsilon =0.5$meV in the two materials, and then for fixed dot radius $a=20$ nm. 
For Si $g=2$, $m^{\ast}=0.2m_{e}$ and $\varepsilon _{r}=12.5$ (Si/SiGe), \cite{Gold,Mori} for
GaAs $g=0.41$, $m^{\ast }=0.067m_{e}$ and $\varepsilon _{r}=12.9$ (GaAs/AlGaAs), \cite{Adachi} where $m_{e}$ is the bare electron mass. 
For $1/f$ noise we assume $S(\omega)$ scales linearly with temperature\cite{Kogan,DuttaHornRMP1981} so we extract the factor $\Delta =\gamma_N k_{B}T$, representing the strength of the noise from Refs. \onlinecite{Takeda} and \onlinecite{Petersson_PRL10} for Si/SiGe and GaAs respectively and scale it to $T=100\,$mK, which gives $\Delta _{\text{Si/SiGe}}=8.85\times 10^{-7}\,$meV$^2$ and $\Delta _{GaAs}=3.79\times 10^{-3}\,$meV$^2$.

\begin{figure}[tbp]
	\caption{Time evolution of the spin in Si/SiGe and GaAs with dot radius $%
	a=20 $nm, cut-off $\protect\omega_0=1$s and all other dot parameters
	identical.}
	\label{figplot}\centering
	\includegraphics[scale=0.9]{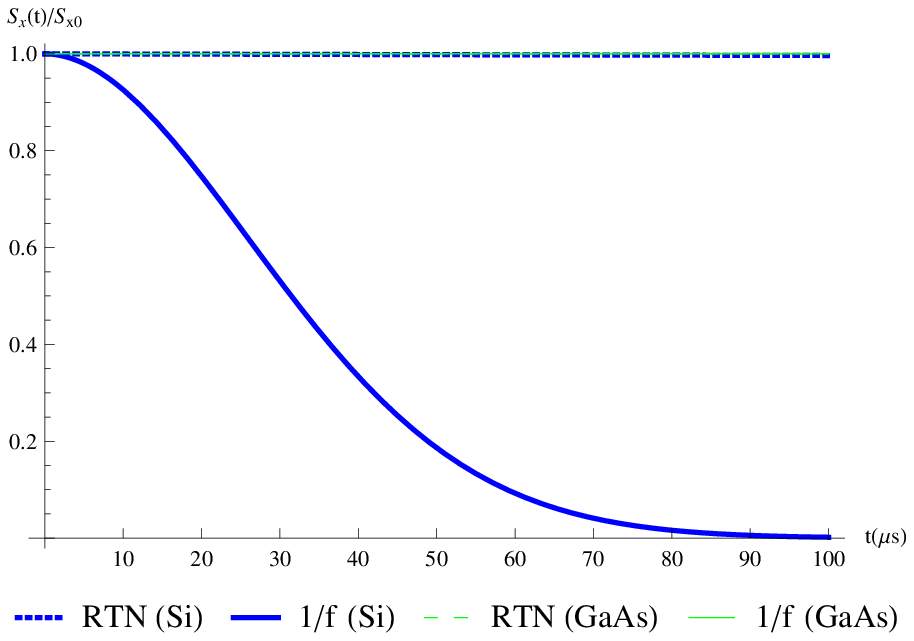} \centering
	\includegraphics[scale=0.9]{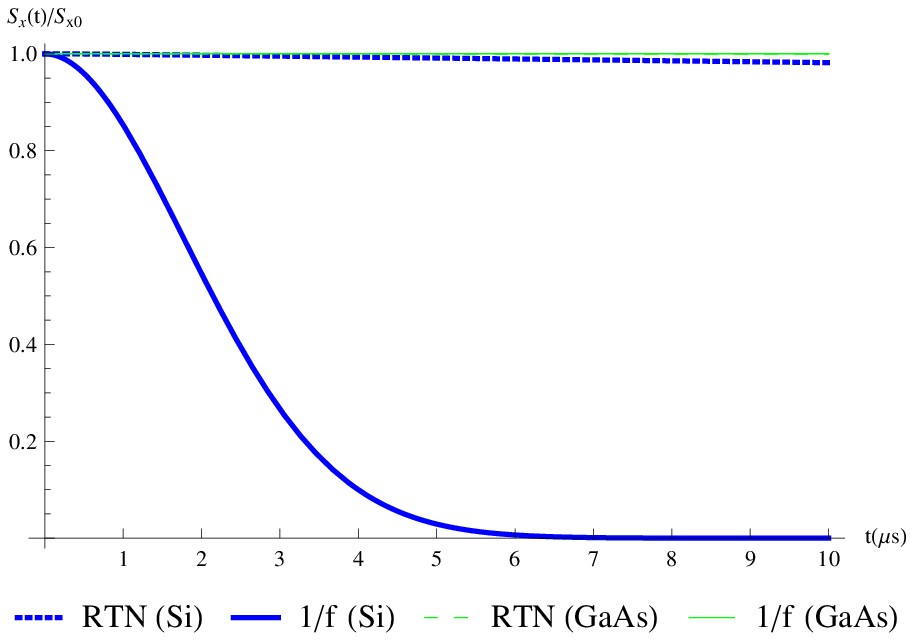}
	\caption{Time evolution of the spin in Si/SiGe and GaAs with fixed
	confinement energy $\protect\delta\protect\varepsilon=0.5$meV, cut-off $%
	\protect\omega_0=1$s and $x_d/a =2$, with all other dot parameters
	identical. }
	\label{figplot2}
	\end{figure}

The results are presented in Tables \ref{resultstable1} and \ref{resultstable2}, which are the main results of this paper, and since Eqs. \ref{eq:T2*RTN} and \ref{eq:T2*1/f} only give estimates for $T_{2}^{\ast }$ we plot the time evolution of the spin $S_{x}(t)$ in Figures \ref{figplot} and \ref{figplot2}. 
We also compare dephasing times with those due to spin-orbit coupling as calculated in Ref. \onlinecite{Berme}. 
In Si/SiGe the dephasing times for spin-orbit and the slanting magnetic field are essentially the same, while in GaAs spin-orbit has a far more destructive effect on qubit coherence compared to the magnetic field. 
Although the numbers in Ref. \onlinecite{Berme} are estimates, the noise profile assumed was the same as in this work. 
The much weaker effect on GaAs is due to the smaller $g$-factor. 
The Overhauser field in GaAs QDs is several orders of magnitude larger than in Si QDs,\cite{Assali_Hyperfine_PRB11} and relevant energy scales have the same order of magnitude as for spin-orbit interactions,\cite{Berme} so without considering feedback mechanisms\cite{Bluhm_Nuclear_Bath_PRL2010} or echo sequences\cite{Bluhm_LongCoh_NP11} designed to increase dephasing times, the contribution from the Overhauser field is the same as due to spin-orbit. 
The contribution of the nuclear spin bath to qubit decoherence for a Si QD can be drastically reduced by using isotopically purified $^{28}$Si samples.\cite{Assali_Hyperfine_PRB11,Muhonen_Store_14}


We also find that $T_2^{\ast}$ is heavily dependent on the QD radius and confinement energy.
$(T_2^{\ast})_{1/f}\propto \delta\varepsilon^4$ and $(T_2^{\ast})_{RTN}\propto \delta\varepsilon^6$, so by doubling the confinement energy, or equivalently reducing the dot radius by a factor of $\sqrt{2}$, the dephasing time can be increased by an order of magnitude. 
Dephasing times can also be increased by reducing the noise spectrum by going to lower temperatures since $S(\omega)\propto T$, or reducing sources of charge noise. 
The latter has recently been achieved by developing quantum dots in an undoped Si/SiGe wafer,\cite{Obata_2014JPSCP} with results indicating that doped materials produce more charge noise sources via the 2DEG and interface trapping sites. 
Moreover, the Rabi oscillation term is $\propto \eta$. 
The dephasing rate for RTN is $\propto\eta^4$, while that due to $1/f$ noise is approximately $\propto \eta^2$.
Reducing $\eta$ therefore reduces the $1/f$ noise dephasing rate faster than the Rabi oscillation gate time (considerably faster for RTN).

	\begin{table}[t]
	\caption{Sample $T_2^*$ for a quantum dot with $a=20\,$nm, defect position $%
	x_d=40\,$nm, $\protect\tau=1\,$ms for RTN, $T=100\,$mK, $\protect\varepsilon%
	_Z \approx 60\,\protect\mu$eV, $S(\protect\omega)$ for $1/f$ noise
	estimated from Refs. \onlinecite{Takeda,Petersson_PRL10}, and $\varepsilon_r = 12.5, 12.9$  for Si/SiGe, GaAs
	respectively. For spin-orbit we used Rashba coefficients from
	Ref. \onlinecite{Berme}.}
	\label{resultstable1}\renewcommand{\arraystretch}{1.3} \centering
	\begin{tabular}{c|c|c|c|c|c|c}
	\hline\hline
	& $\delta\varepsilon$ & RTN + $\mathbf{B}$ & RTN+SO & $1/f+\mathbf{B}$ & $%
	1/f+\text{SO}$ &  \\ \hline
	Si/SiGe & 1 meV & 30 ms & 1 ms & 130 $\mu$s & 30 $\mu$s &  \\ 
	GaAs & 3 meV & 1900 s & 25 ns & 7 s & 25 $\mu$s &  \\ \hline\hline
	\end{tabular}%
	\par
	\renewcommand{\arraystretch}{1.3} \centering
	\begin{tabular}{c|c|c|c|c|c|c}
	\hline\hline
	& $a$ & RTN + $\mathbf{B}$ & RTN+SO & $1/f+\mathbf{B}$ & $1/f+\text{SO}$ & 
	\\ \hline
	Si/SiGe & 30 nm & 520 $\mu$s & 310 $\mu$s & 10 $\mu$s & 7 $\mu$s &  \\ 
	GaAs & 50 nm & 40 ms & 500 ps & 7 ms & 770 ns &  \\ 
	GaAs$^\dagger$ & 50 nm & 610 $\mu$s & 500 ps & 840 $\mu$s & 770 ns &  \\ 
	\hline\hline
	\end{tabular}%
	\caption{Sample $T_2^{\ast}$ for a quantum dot with confinement energy $%
	\protect\delta\protect\varepsilon=0.5\,$meV, and all parameters except $%
	a,x_d $ identical to Table \protect\ref{resultstable1}; we fix $x_d/a =2
	$. $^\dagger$We set $\protect\eta$ to be constant, so for GaAs, $b_{sl}$ is
	$\sim 3$ times larger than for Si to account for the smaller $g$-factor (not
	5 times larger as $\protect\eta\propto a$ also.)}
	\label{resultstable2}
	\end{table}

Inhomogeneous magnetic fields are also an essential ingredient of singlet-triplet qubits.\cite{Wu2014, Wong2014,Thalineau_PRB2014, Shulman_Science, Maune_Nature2012} 
There, however, dephasing will be dominated by direct coupling to charge noise via exchange. \cite{Culcer_APL09} 
The slanting magnetic field yields a term $\propto \tilde{\sigma}_{x} $, which gives relaxation, but not dephasing.

\textit{Evanescent-wave Johnson noise from a micromagnet}. 
For an electron spin qubit in a fluctuating magnetic field 
	\begin{equation*}
	\frac{1}{T_{1}}=\frac{\mu _{B}^{2}}{\hbar ^{2}}\sum_{i=x,y}\left\langle
	B_{i}\left( \mathbf{r}\right) B_{i}\left( \mathbf{r}\right) \right\rangle
	_{\omega _{0}},
	\end{equation*}
where $\mathbf{r}$ is the position of the qubit and $\omega_{0}$ is the Zeeman splitting caused by the applied field $\mathbf{B}_{0}$ taken here to be $\parallel\hat{\bm{z}}$. 
Only the transverse components of $\mathbf{B}$ contribute to $T_{1}$. 
If a micromagnet is placed at the origin of the coordinates, then, with $\mathbf{r}=\left( x_{1},x_{2},x_{3}\right) $, \cite{Supplement} 
	\begin{IEEEeqnarray}{rCl}
	\langle B_{i}  B_{j} \rangle_{\omega_{0}}= \hbar V\text{ Im }\alpha\left( \omega_{0}\right) \coth \left( \frac{\hbar \omega_{0}}{2k_{B}T}\right) \frac{3x_{i}x_{j}+\delta_{ij}r^{2}}{r^{8}} \nonumber\\
	\label{eq:bnoise}
	\end{IEEEeqnarray}
Considering only finite frequencies, the magnetic polarizability $\alpha \left( \omega \right) $ is defined by $\mathbf{m} \left( \omega \right) =V\alpha \left( \omega \right) \mathbf{h}\left( \omega
\right) $, where $\mathbf{m}\left( \omega \right) $ is the total magnetic moment of the micromagnet particle, $V$ is the volume of the magnet, and $\mathbf{h}$ is a spatially uniform applied magnetic field.

We need to estimate the dissipative (imaginary) part of $\alpha$. In general $\alpha $ is a matrix, but we are mainly interested in order-of-magnitude estimates, so we ignore anisotropies and factors of order unity. This is justified below. The dynamics of the magnet are dominated by ferromagnetic resonance. In principle, this is dangerous for qubit decoherence in Si ($g\approx 2)$, since the fundamental resonance frequency $g\mu _{B}B_{0}/\hbar $ for the precessions of the qubit spin and the \textit{bare} resonance frequency $\omega _{0}$ of the Co magnet ($g$-factor also very close to 2) are close.

Let $\mathbf{M}_{0}$ be the permanent magnetization and let this be $\parallel\mathbf{B}_{0}$, since this is the geometry chosen in all experiments to date. 
Following the established theory of ferromagnetic resonance, \cite{Vonsovskii} we decompose the magnetization and field into a large time-independent term and a small time-dependent term. 
Thus we have $\mathbf{M}\left( t\right) =\mathbf{M}_{0}+\mathbf{m}\left( t\right) $ and $\mathbf{H}
=\mathbf{B}_{0}+\mathbf{h}\left( t\right) $. 
The equation of motion neglecting spin-orbit coupling and shape anisotropy is: 
	\begin{equation*}
	\frac{d\mathbf{M}}{dt}=\frac{d\mathbf{m}}{dt}\approx -\gamma \left( \mathbf{M}_{0}\times \mathbf{h}+\mathbf{m}\times \mathbf{B}_{0}\right) .
	\end{equation*}	
The first term is the change in the macroscopic $\mathbf{M}$ that is produced by $\mathbf{h},$ and the second term is the precession of $\mathbf{m}$ in the external field. 
Changing to the frequency domain at a fixed driving frequency $\omega $ leads to 
	\begin{equation*}
	-i\omega \mathbf{m}=\gamma (M_{0}\mathbf{h}-B_{0}\mathbf{m})\times \hat{z}
	=(\omega _{M}\mathbf{h}-\omega _{0}\mathbf{m})\times \hat{z}
	\end{equation*}
where $\omega $ is the driving frequency: $\mathbf{h}\left( t\right) = \mathbf{h}e^{-i\omega t},$ $\omega _{0}=\gamma B_{0}$ is the "bare" Larmor frequency associated with the DC applied field $\mathbf{B}_{0}$ and $\omega_{M}$ is the frequency associated with $\mathbf{M}_{0}:$ $\omega _{M}=\gamma M_{0};$ $\gamma =g\mu _{B}/\hbar $. 
Solving these equations we find $\mathbf{m}=\chi \mathbf{h}+iG\mathbf{h}\times \hat{z}$, with $\chi \left( \omega \right) =\omega _{M}\omega _{0}/\left( \omega _{0}^{2}-\omega ^{2}\right) $ and $G\left( \omega \right) =\omega _{M}\omega /\left( \omega_{0}^{2}-\omega ^{2}\right) $. 
Here $\chi $ acts as an ordinary (diagonal) susceptibility while the $G$ term gives the effect of precession about the magnetization. 
We will not compute the effects of spin-orbit coupling and shape anisotropy in detail, since they depend on such unknowns as the microcrystallinity of the magnet. 
We only note that the resonance frequency $\omega _{0}$ of the magnet is strongly shifted by the restoring forces due to these effects. 
Hence the resonance frequencies of the magnet and qubit, though of the same order of magnitude, in general do not coincide, and $\omega _{0}$ should be replaced by the physical resonance frequency $\omega _{res}\neq \omega _{0}$. 
This physical picture is in good agreement with the noise spectra in Ref. \onlinecite{Neumann_JAP2015}, in which nearly all the weight is
shifted upwards from the bare resonance frequency.

So far all the response is real and there is no dissipation. 
The form of the damping term depends on the mechanism. 
However, at the phenomenological level of the present treatment, a simple Bloch-Bloembergen form
	\begin{IEEEeqnarray*}{rCl}
	\frac{d{\bm M}}{dt} &=&-\gamma \mathbf{M}\times \mathbf{H} - \bigg(\frac{{\bm M} - M_0 \hat{\bm z}}{T_{2}}\bigg),
	\end{IEEEeqnarray*}
is sufficient, which leads to the final result 
	\begin{IEEEeqnarray}{rCl}
	\text{Im}\chi \left( \omega \right) =\omega _{M}\omega _{res}~\frac{2\omega	T_{2}^{-1}}{\left( \omega _{res}^{2}-\omega^{2}+T_{2}^{-2}\right)^{2}+4T_{2}^{-4}},  \notag
	\end{IEEEeqnarray}
and similarly for $\text{Im }G$. 
There are too many unknowns to attempt a very accurate estimate of $\chi $ and $G$ but we may
use values of parameters from bulk Co\cite{Frait} and Co wires\cite{FerrePhysRevB.56.14066,RespaudPhysRevB.59.R3934} to make an order-of-magnitude estimate.

It is simplest to estimate the effect of the noise in the experiment of Ref. \onlinecite{PhysRevLett.96.047202}, in which the separation $\mathbf{r}$ from the magnet to the qubit satisfies $\left\vert \mathbf{r}\right\vert \gg L,$ where $L$ is the largest linear dimension of the magnet.
This is not the case in the set-ups in Refs. \onlinecite{Nowack_Science07,Pioro-L2,Pioro-L1}, and \onlinecite{Kawakami2014}, which has larger magnets closer to the qubit. 
For an analysis of this type of device, see Ref. \onlinecite{Neumann_JAP2015}.
Taking $B_{0}=20$ mT from Ref. \onlinecite{PhysRevLett.96.047202} as the applied field, we find $\omega _{M}=28$ GHz using the formulas above and from experiment $T_{2}$(magnet) $\approx 10^{-9}$ s. \cite{Frait} 
$\omega_{res}$ has the same order of magnitude as $\omega _{0}=4$ GHz. 
Since all three quantities that enter $\text{Im}\chi $ are roughly of the same order, $\text{Im}\chi $ and $\text{Im}\alpha $ can be taken to be of order unity, and indeed $\alpha $ cannot much exceed unity since it is proportional to the fraction of energy absorbed from the time-dependent magnetic field $\mathbf{h}\left( t\right)$. 
Focusing on the design in Ref. \onlinecite{Wu2014} (somewhat different from Fig. \ref{figslanting}), we substitute $r=1.8$ $\mu $m and find $T_{1}\approx T_{2}^{\ast }\approx 10$ s. (Since the Johnson noise is white, the longitudinal and transverse decoherence times do not usually differ by a factor of 2.) 
This value for $T_{2}^{\ast },$ is far longer than the measured $T_{2}^{\ast} \approx 200$ ns, \cite{PhysRevLett.96.047202} implying that the micromagnet noise is not contributing to decoherence in this experiment. 
Ref. \onlinecite{Neumann_JAP2015} also found $T_{1}\gtrsim 10$ s, which suggests that when $\left\vert \mathbf{r}\right\vert \sim L$ there is some cancellation in the noise fields.

In conclusion, we have studied the contribution to dephasing of an applied inhomogeneous magnetic field and charge noise on a single-spin qubit and found it is an effective source of qubit decoherence particularly for Si/SiGe devices. 
Our results imply that when implementing slanting magnetic field architectures for spin control, noise sources need to be considered and reduced to improve coherence times. 
By contrast, the Johnson noise from the micromagnet is probably not a significant source of decoherence in the current generation of experiments. 
It may become important as decoherence times become longer, of the order of seconds.

We are grateful to L. Vandersypen for discussions, Andrea Morello and M. Pioro-Ladri\`{e}re for bringing to our attention existing micromagnet designs, and to M. A. Eriksson for discussion of experimental devices. 
We would also like to acknowledge support from the Gordon Godfrey bequest. 
This research was partially supported by the US Army Research Office (W911NF-12-0607).

\appendix

\section{Total Hamiltonian}

The total Hamiltonian is $H = H_0+H_N+H_Z$, which, in the basis $\{D_{0,\uparrow}, D_{0,\downarrow}, D_{+,\uparrow}, D_{+,\downarrow}, D_{-,\uparrow}, D_{-,\downarrow}\}$, is written as 
	\begin{equation}
	\tag{S2}
	H=
	\label{eq:H}
	\begin{pmatrix}
	\begin{tabular}{c c | c c c c}
	$\varepsilon_0^+$ & $0$ & $\alpha$ & $\eta$ & $\alpha$ & $\eta$\\
	$0$ & $\varepsilon_0^-$ & $\eta$ & $\alpha$ & $\eta$ & $\alpha$\\
	\hline
	$\alpha$ & $\eta$ & $\varepsilon_1^+$ & $0$ & $\beta$ & $0$\\
	$\eta$ & $\alpha$ & $0$ & $\varepsilon_1^-$ & $0$ & $\beta$\\
	$\alpha$ & $\eta$ & $\beta$ & $0$ & $\varepsilon_1^+$ & $0$\\
	$\eta$ & $\alpha$ & $0$ & $\beta$ & $0$ & $\varepsilon_1^-$
	\end{tabular}
	\end{pmatrix},
	\end{equation}	
where $\varepsilon_0^\pm=\varepsilon_0+v_0\pm\varepsilon_Z$ and $\varepsilon_1^\pm=\varepsilon_1+v_1\pm\varepsilon_Z$ are the Zeeman-split orbital levels including the charge noise and magnetic field contributions, and the magnetic field matrix elements $\varepsilon_Z = \langle D_{0,\uparrow}|H_Z|D_{0,\uparrow}\rangle$ and $\eta=\tfrac{1}{2}g\mu_B\langle D_0|B_z|D_+\rangle=\tfrac{1}{2}g\mu_B\langle D_-|B_z|D_0\rangle$. 

In the Schrieffer-Wolff transformation we keep first order terms in $\delta v(t)$ and $\varepsilon_z$, and neglected terms proportional to the identity matrix and non-fluctuating terms (terms not involving $\delta v(t)$). 

\section{Formal treatment of dephasing}

The qubit is described by a spin density matrix $\rho(t)=\tfrac{1}{2}\boldsymbol{\sigma}\cdot\mathbf{S}(t)$ which satisfies the quantum Liouville equation $d\rho/dt + (i/\hbar)[H,\rho]=0$.
The fluctuating $z$-component of the effective Hamiltonian causes dephasing, which for RTN is $\mathbf{V}(t)=V(t)\sigma_z(-1)^{N(t)}$, where $N(t)=0,1$ is a Poisson random variable with fluctuator switching time $\tau$. 
We work in a rotating reference frame which takes into account the effect of the laboratory effective magnetic field, assumed to be spatially homogeneous, in which the $z$-component of the spin projection is conserved so we are studying pure dephasing. 
To determine the full time evolution of the density matrix with initial conditions $\rho(0)=\frac{1}{2}\boldsymbol{\sigma}\cdot\boldsymbol{S}_0$, we
define $\mathbf{h}(t)\equiv\frac{1}{\hbar}\int_{0}^{t}\boldsymbol{V}(t^{\prime})dt^{\prime}$, with $h(t)=|\mathbf{h}(t)|$. 
The time evolution of the $i^{\text{th}}$ spin component $S_i(t)=tr[\sigma_i\rho(t)]$ is
	\begin{equation}
	  \makebox[0pt]{\begin{minipage}{\linewidth}
		\begin{IEEEeqnarray}{rCl}
		S_i(t)&=& S_{0i}\cos{h(t)}-\epsilon_{ijk}\hat{h}_k(t)\sin{h(t)}\nonumber\\
		&+&\hat{h}_i(t)[\boldsymbol{\hat{h}}(t)\cdot\boldsymbol{S}_0][1-\cos{h(t)}].\nonumber
		\end{IEEEeqnarray}
	  \end{minipage}}
	  \tag{S3}
	\end{equation}
If $S_x(0)=S_{0x}\boldsymbol{\hat{x}}$ then $S_x(t)\approx S_{0x}\cos{h(t)}$, and taking the average over the realisations of $\cos{h(t)}$,\cite{PhysRevB.68.115322,Culcer,Culcer1,Berme} 
	\begin{equation}
	\langle\cos{h(t)}\rangle= e^{-t/\tau} \Bigg[\frac{\sinh{t}}{\Xi\tau}+\cosh{\Xi t}\Bigg],
	\label{eq:avgcosh}
	\tag{S4}
	\end{equation}
where $\Xi = \sqrt{(1/\tau)^2 - V^2/\hbar^2}$. We consider fluctuators with switching times $\tau<1\,\mu$s, in which the condition $V^2\ll(\frac{\hbar}{\tau})^2$ is satisfied and we may expand Eq. \eqref{eq:avgcosh} in $\frac{V^2\tau^2}{\hbar^2}$. The initial spin decays exponentially as $S_x(t)=S_{0x}e^{-t/T_2^{\ast}}$, with 
	\begin{equation}
	\tag{S5}
	\label{eq:T2*RTN}
	\left( \frac{1}{T_2^{\ast}}\right)_{\text{RTN}}=\frac{V^2\tau}{2\hbar^2},
	\end{equation}
where for the slanting magnetic field $V=\frac{8\eta^2\delta v(t)\varepsilon_z}{\delta\varepsilon^3}$.

For $1/f$ noise the main contribution is concentrated at low frequencies, so it primarily affects qubit dephasing in both semiconductor\cite{PhysRevB.73.180201,Paladino_1/f_RMP14,Burkard_Decoh_AdvPhys08,Culcer_APL09} and superconductor\cite{Clarke_Nat2008,Makhlin_RMP2001} devices. It is typically Gaussian in semiconductors\cite{Kogan} and can be described by the correlation function $S(t)=\langle V(0)V(t)\rangle$. The Fourier transform of this is the noise spectral density which has the form $S(\omega)=\frac{\gamma_N k_BT}{\omega}$, where $\gamma_N$ (units of energy) is obtained from experiment. Hence, for $\mathbf{V}(t)$ in the qubit subspace, we have $S_V(\omega)\approx\left(\frac{8\eta^2\varepsilon_z}{\delta\varepsilon^3}\right)^2 S(\omega)$. 
We write $S_x(t)=S_{0x}e^{-\chi(t)} $, where \cite{deSousa}
	\begin{equation}
	\tag{S6}
	\chi(t)= \frac{2\gamma_N k_B T}{\hbar^2}\left(\frac{8\eta^2\varepsilon_z}{\delta\varepsilon^3}\right)^2\int_{\omega_0}^{\infty}d\omega\left(\frac{\sin^2\omega t/2}{\omega^3}\right),
	\end{equation}
for a low-frequency cut-off $\omega_0$ typically chosen as the inverse of the measurement time. 
We assume $\omega_0t\ll 1$ and we can approximate $\chi(t)\approx\left( \frac{t}{T_2^{\ast}}\right)^2\ln{\frac{1}{\omega_0t}}$, with 
	\begin{equation}
	\tag{S7}
	\label{eq:T2*1/f}
	\left( \frac{1}{T_2^{\ast}}\right)_{1/f}\approx \sqrt{\frac{\gamma_N k_BT}{2\hbar^2}}\left(\frac{8\eta^2\varepsilon_z}{\delta\varepsilon^3}\right).
	\end{equation}

\section{Johnson Noise}
The detailed derivation of Eq. (6) will be given elsewhere. However, its qualitative form is not difficult to understand by analogy with an interaction of the van der Waals type.\cite{LLv9} Let the ferromagnetic particle be located at the origin. A momentary fluctuation of the magnetic dipole of the qubit produces a corresponding fluctuation of the magnetic field $\sim 1/r^{3}$ at the ferromagnet. This induces a magnetic polarization $\mu \sim\alpha /r^{3}$ of the magnet which in turn causes a field $B\sim\mu /r^{3} \sim\alpha /r^{6}$ at the qubit. 
The temperature dependence is specified by the fluctuation-dissipation theorem, which also requires that it is only the dissipative part of $\alpha$ that contributes.


%

\end{document}